\documentclass[pra,aps,twocolumn,floatfix,showpacs]{revtex4}
\usepackage{graphicx,psfrag,amsmath,amssymb,times}

\begin{document}
\title{Dimer-atom scattering between two identical fermions and a third particle}
\author{M. Iskin}
\affiliation{Department of Physics, Ko\c c University, Rumelifeneri Yolu, 34450 Sariyer, Istanbul, Turkey}
\date{\today}

\begin{abstract}
We use the diagrammatic $T$-matrix approach to analyze the three-body 
scattering problem between two identical fermions and a third 
particle (which could be a different species of fermion or a boson). 
We calculate the s-wave dimer-atom scattering length 
for all mass ratios, and our results exactly match the 
results of Petrov. In particular, we list the exact dimer-atom 
scattering lengths for all available two-species Fermi-Fermi and 
Bose-Fermi mixtures. In addition, unlike that of the equal-mass 
particles case where the three-body scattering $T$-matrix 
decays monotonically as a function of the outgoing momentum, 
we show that, after an initial rapid drop, this function 
changes sign and becomes negative at large momenta and then decays 
slowly to zero when the mass ratio of the fermions to the third particle
is higher than a critical value (around $6.5$). As the mass ratio 
gets higher, modulations of the $T$-matrix become more apparent 
with multiple sign changes, related to the ``fall of a particle 
to the center'' phenomenon and to the emergence of three-body 
Efimov bound states.
\end{abstract}
\pacs{03.75.Ss, 03.75.Hh, 05.30.Fk}
\maketitle

\section{Introduction}
\label{introduction}

The dimer-atom scattering process was first solved by Skorniakov 
and Ter-Martirosian in 1956~\cite{skorniakov}, in the context of 
three-nucleon scattering, i.e. a two-body bound state between a 
neutron and a proton (called a deuteron) is scattering with a neutron. 
They considered equal-mass particles, and found that while the 
Born approximation gives $a_{DA} = 8a_{AB}/3$ for the scattering 
length between an AB dimer and an A atom, where $a_{AB}$ is the 
two-body scattering length between A and B particles, the exact 
result turns out to be $a_{DA} \approx 1.18a_{AB}$~\cite{skorniakov}.
This problem has recently been analyzed with the diagrammatic 
$T$-matrix approach~\cite{brodsky, levinsen}, in the context 
of cold quantum gases (see also~\cite{bedaque98} in a different 
context), and the result is in perfect agreement with the 
earlier results. In the same context, the three-body 
scattering problem has also been generalized to unequal-mass 
particles, and analyzed both in real space through solving 
the three-body Schr\"odinger equation~\cite{petrov-abf} 
and in momentum space through the diagrammatic $T$-matrix 
approach~\cite{iskin-bf}, with perfect agreement in between for all 
mass ratios.

When the mass ratio of the heavy particles to the light one 
is below $13.61$, it is also well-established that the three-body 
scattering problem is universal, and that the dimer-atom 
scattering length is proportional to the two-body scattering length $a_{AB}$ 
(just like the equal mass case) with the proportionality factor depending 
only on the masses of the constituent particles~\cite{petrov-abf}. 
However, since three-body Efimov bound states emerges for larger mass 
ratios~\cite{efimov, petrov, kartavtsev}, this problem is not universal, 
and an additional parameter coming from the short-range (or large-momentum) 
three-body physics is needed for an accurate description.

Here, we use the diagrammatic $T$-matrix approach to analyze 
the three-body scattering problem between two identical fermions and 
a third particle. Our results are relevant to the quantum phases of 
two-species Fermi-Fermi~\cite{taglieber08,wille08,voigt09,spiegelhalder09,tiecke09,spiegelhalder10}. 
and Bose-Fermi~\cite{inouye04, ospelkaus06, zaccanti, zirbel} 
mixtures of atomic gases in the molecular limit. 
The remaining paper is organized as follows. 
After deriving the Skorniakov and Ter-Martirosian integral equation 
generalized for unequal-mass particles in Sec.~\ref{sec:t-matrix}, 
we numerically solve the resultant equation for all mass ratios 
in Sec.~\ref{sec:a_DA}, and list the exact dimer-atom scattering 
lengths for all available two-species Fermi-Fermi and 
Bose-Fermi mixtures in Sec.~\ref{sec:mixtures}. A brief summary 
of our conclusions is given in Sec.~\ref{sec:conclusions}.

\section{Three-body problem}
\label{sec:three-body}

In this paper, we are interested in the three-body (dimer-atom) 
s-wave scattering between two identical fermions (refer to it as 
A-type particles) and a third particle (refer to it as a B-type 
particle which could be a different species of fermion or a boson). 
In particular, we consider a zero-ranged attractive interaction 
between A and B particles, and assume there is a weakly bound resonance 
between them with the binding energy $\epsilon_b < 0$, so that 
we want to study the scattering between this bound state 
(refer to it as an AB dimer) and the remaining A atom.

\subsection{Dimer-atom scattering $T$-matrix}
\label{sec:t-matrix}

Detailed description of the diagrammatic $T$-matrix approach for
the dimer-atom scattering process can be found in the literature for
equal-mass particles~\cite{brodsky, levinsen, bedaque98}, 
and here we give details of our calculation~\cite{iskin-bf} for 
the case of unequal-mass particles.

\begin{figure} [htb]
\psfrag{a}{\LARGE $-\mathbf{k}, \omega_A$}
\psfrag{b}{\LARGE $\mathbf{k}, \omega_D + \epsilon_b$}
\psfrag{c}{\LARGE $\mathbf{k}+\mathbf{p}, \omega_D - \omega_A + \epsilon_b + p_0$}
\psfrag{d}{\LARGE $-\mathbf{q}, \omega_A - q_0$}
\psfrag{i}{\LARGE $\mathbf{q}, \omega_D + \epsilon_b + q_0$}
\psfrag{f}{\LARGE $\mathbf{q}+\mathbf{p}, \omega_D - \omega_A + \epsilon_b + p_0 + q_0$}
\psfrag{g}{\LARGE $-\mathbf{p}, \omega_A - p_0$}
\psfrag{h}{\LARGE $\mathbf{p}, \omega_D + \epsilon_b + p_0$}
\centerline{\scalebox{0.4}{\includegraphics{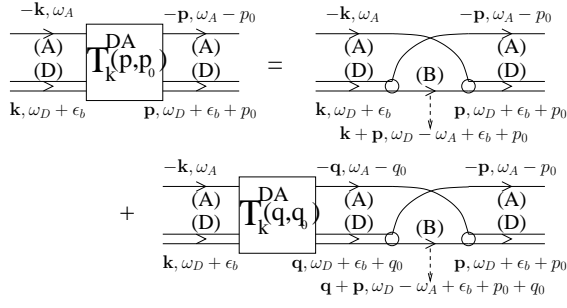}}}
\caption{\label{fig:t-matrix}
Diagrammatic representation of the integral equation for the
three-body scattering $T$-matrix $T_\mathbf{k}^{DA}(\mathbf{p},p_0)$ 
between an AB dimer (dimer D consists of one fermionic particle A and a 
second particle B) and a fermionic particle A. Particle B 
could be a different species of fermion or a boson.
}
\end{figure}

We begin our analysis by describing the zero temperature diagrammatic 
representation for the dimer-atom scattering $T$-matrix 
$T_\mathbf{k}^{DA}(\mathbf{p},p_0)$ as illustrated in Fig.~\ref{fig:t-matrix},
where $\mathbf{k}$ and $\mathbf{p}$ are the incoming and outgoing momenta, 
and $\omega_A = k^2/(2m_A)$ and $\omega_D = k^2/(2m_D)$ are the 
frequencies for the incoming A particle and AB dimer, respectively.
Here, we set the center-of-mass momentum of the dimer-atom system to 
zero, and $\epsilon = \omega_A + \omega_B + \epsilon_b$ is the total
frequency where $\epsilon_b = - 1/(m_{AB} a_{AB}^2) < 0$ is the binding 
energy of the two-body bound state between A and B particles,
and $m_D = m_A + m_B$ and $m_{AB} = 2 m_B m_A/m_D$
are masses of the AB dimer and twice the reduced mass of 
the $A$ and $B$ particles, respectively.

In Fig.~\ref{fig:t-matrix}, single lines represent retarded free 
propagators for the A and B particles
\begin{eqnarray}
\label{eqn:atom-propagators}
G_{A,B}(\mathbf{k}, \omega) &= \frac{1}{\omega - \omega_{A,B} + \mu_{A,B} + i0^+},
\end{eqnarray}
where $\omega_{A,B} = k^2/(2m_{A,B})$ is the energy, and $\mu_{A,B}$ 
is the chemical potential of the corresponding particle.
Similarly, double lines represent the retarded propagator for the 
AB dimer, and in three dimensions it can be approximated by a simple 
one-pole structure
\begin{equation}
\label{eqn:dimer-propagator}
G_D(\mathbf{k}, \omega) = \frac{\frac{4\pi}{m_{AB}}}
{\frac{1}{a_{AB}} - \left[m_{AB}(\omega_D - \omega - \mu_A - \mu_B - i0^+)\right]^{1/2}},
\end{equation}
which reflects the presence of a two-body bound state between 
A and B particles. 

This dimer propagator is obtained from the resummation of the 
AB polarization bubbles leading to
$
G_D(\mathbf{k}, \omega) =  - g/[1 + g \Gamma_{AB} (\mathbf{k}, \omega)],
$
where $g > 0$ is the strength of the bare interaction between A and
B particles, and the AB polarization bubble is 
$
\Gamma_{AB} (\mathbf{k}, \omega) = \sum_{\mathbf{q}, q_0} G_A(\mathbf{k} + \mathbf{q}, \omega + q_0 ) 
G_B(-\mathbf{q}, -q_0 ).
$
At zero temperature, $\sum_{\mathbf{q}, q_0} \equiv i\int d\mathbf{q} dq_0/(2\pi)^4$
in three dimensions. Integration over the internal momentum 
$\mathbf{q}$ and frequency $q_0$ leads to 
$
\Gamma_{AB} (\mathbf{k}, \omega) = \Gamma_{AB} (\mathbf{0},0)
+ m_{AB}^{3/2} \left( \omega_D - \omega  - \mu_A - \mu_B - i0^+ \right)^{1/2} / (4 \pi),
$
which in combination with the definition of the two-body scattering 
length $a_{AB} = m_{AB} T^{AB} (\mathbf{0},0)/ (4\pi)$, 
and the two-body scattering $T$-matrix 
$
T^{AB} (\mathbf{0},0) = - g / [1 + g \Gamma_{AB} (\mathbf{0}, 0)],
$
between A and B particles, give the final result described 
in Eq.~(\ref{eqn:dimer-propagator}).

In the following, we set $\mu_A = \mu_B = 0$ since we are interested
in the dimer-atom scattering in vacuum. However, note that our diagrammatic 
calculation for the scattering parameters of the three-body 
problem is exact, and they are sufficient to describe ultracold quantum 
gases, since experiments are always performed at low densities. The 
calculation of scattering parameters in the presence of many other particles
(arbitrary density) is much more difficult, and it is not discussed here. 
In addition, on the right hand side of Fig.~\ref{fig:t-matrix}, note 
that the first diagram represents a fermion exchange process, 
i.e. first the particle B breaks up with the particle A and then it 
forms a new AB dimer with the remaining A particle. This is the 
simplest process contributing to the dimer-atom scattering, 
e.g. Born approximation, and all other (infinitely many) possible 
processes are included in the second diagram. 

In analytical form, the dimer-atom $T$-matrix $T_\mathbf{k}^{DA}(\mathbf{p},p_0)$ 
satisfies the following integral equation
\begin{eqnarray}
\label{eqn:t-matrix}
&& T_\mathbf{k}^{DA}(\mathbf{p},p_0) = 
- G_B(\mathbf{k}+\mathbf{p}, \omega_D - \omega_A + \epsilon_b + p_0) \nonumber \\
&& - \sum_{\mathbf{q}, q_0} G_D(\mathbf{q}, \omega_D + \epsilon_b + q_0) 
G_A(-\mathbf{q}, \omega_A - q_0) \times \\
&& T_\mathbf{k}^{DA}(\mathbf{q},q_0) G_B(\mathbf{p} + \mathbf{q}, \omega_D - \omega_A + \epsilon_b + p_0 + q_0) \nonumber.
\end{eqnarray}
The minus signs on the right hand side is due to Fermi-Dirac statistics, 
i.e. exchanging a fermion brings a minus sign unlike that of a boson. 
The integration over frequency $q_0$ can be 
easily performed by closing the integration contour in the upper 
half-plane, where both $T_\mathbf{k}^{DA}(\mathbf{q},q_0)$ and 
$G_D(\mathbf{q}, \omega_D + \epsilon_b + q_0)$ are analytic functions 
of $q_0$, and only a simple pole contribution comes from 
$G_A(-\mathbf{q}, \omega_A - q_0)$. Note that this property of 
$T_\mathbf{k}^{DA}(\mathbf{q},q_0)$ is due to the form of Eq.~(\ref{eqn:t-matrix}) 
itself. This integration sets $q_0 = (k^2-q^2)/(2m_A)$, and we set 
$p_0 = (k^2-p^2)/(2m_A)$ in order to have the same frequency dependence 
for the $T$-matrix on both sides of Eq.~(\ref{eqn:t-matrix}).
This leads to
\begin{widetext}
\begin{eqnarray}
\label{eqn:t-gen}
T_\mathbf{k}^{DA}(\mathbf{p}) = \frac{m_{AB}}{p^2 
+ \frac{m_{AB}}{m_B}\mathbf{p}\cdot\mathbf{k} + k^2 - m_{AB}\epsilon}
+ \sum_{\mathbf{q}} \frac{4\pi T_\mathbf{k}^{DA}(\mathbf{q})}
{\left(q^2 + \frac{m_{AB}}{m_B}\mathbf{p}\cdot\mathbf{q} + p^2 - m_{AB}\epsilon\right)
\left[\frac{1}{a_{AB}} - \left(\frac{m_{AB}}{m_{DA}} q^2 - m_{AB}\epsilon\right)^{1/2}\right]},
\end{eqnarray}
where we redefine the $T$-matrix $T_\mathbf{k}^{DA}(\mathbf{p}) 
= T_\mathbf{k}^{DA}(\mathbf{p}, (k^2-p^2)/(2m_A))$, and
$\epsilon = k^2/(2m_{DA}) + \epsilon_b$ is the total energy
and $m_{DA} = 2 m_D m_A/(m_D + m_A)$ is twice the reduced mass of 
an AB dimer and an A particle. Since we are interested in the zero-ranged 
low-energy s-wave scattering, we first average out directions of 
the incoming momentum $\mathbf{k}$, and then of the outgoing momentum 
$\mathbf{p}$, leading to
\begin{eqnarray}
T_k^{DA}(p) = \frac{m_B}{2pk} 
\ln\left(\frac{p^2 + \frac{m_{AB}}{m_B}pk + k^2 - m_{AB}\epsilon}
{p^2 - \frac{m_{AB}}{m_B}pk + k^2 - m_{AB}\epsilon}\right)
+ \int_0^\infty \frac{dq \frac{m_{B}}{m_{AB}} \frac{q}{\pi p} T_k^{DA}(q)}
{\frac{1}{a_{AB}} - \left(\frac{m_{AB}}{m_{DA}} q^2 - m_{AB}\epsilon\right)^{1/2}}
\ln\left(\frac{q^2 + \frac{m_{AB}}{m_B}pq + p^2 - m_{AB}\epsilon}
{q^2 - \frac{m_{AB}}{m_B}pq + p^2 - m_{AB}\epsilon}\right),
\end{eqnarray}
for three-dimensional systems, where 
$
T_k^{DA}(p) = \int d\Omega_\mathbf{p} \int d\Omega_\mathbf{k} 
T_\mathbf{k}^{DA}(\mathbf{p}) / (4\pi)^2
$
is the angular-averaged $T$-matrix. 

To obtain the three-body (dimer-atom) s-wave scattering length $a_{DA}$, 
the total energy $\epsilon$ should be set to the binding energy $\epsilon_b$ 
of the two-body bound state, in the limit of vanishing incoming
and outgoing momentum and frequency, i.e. $a_{DA} = 2m_{DA} T_0^{DA}(0)/m_{AB}^2$.
This motivates us to introduce the dimer-atom scattering function 
$a_{k}^{DA} (p)$ for which the integral equation becomes
\begin{eqnarray}
\frac{\frac{m_{AB}}{m_{DA}} a_0^{DA}(p)} 
{\frac{1}{a_{AB}} + \left(\frac{m_{AB}}{m_{DA}} p^2 + \frac{1}{a_{AB}^2}\right)^{1/2}} 
= \frac{1}{p^2 + \frac{1}{a_{AB}^2}} - \frac{m_B}{\pi p m_{AB}}
\int_0^\infty \frac{dq}{q} 
\ln\left( \frac{q^2 + \frac{m_{AB}}{m_B}qp + p^2 + \frac{1}{a_{AB}^2}} 
{q^2 - \frac{m_{AB}}{m_B}qp + p^2 + \frac{1}{a_{AB}^2}} \right) a_0^{DA}(q),
\label{eqn:ieqn}
\end{eqnarray}
\end{widetext}
in the limit when $k \to 0$. Here, the dimer-atom scattering length is 
$a_{DA} = a_0^{DA}(0)$, where
\begin{equation}
a_k^{DA}(p) = \frac{m_{DA}}{m_{AB}^2}
\left[\frac{1}{a_{AB}} + \left(\frac{m_{AB}}{m_{DA}} p^2 - m_{AB}\epsilon\right)^{1/2} \right] T_k^{DA}(p),
\label{eqn:a-scattering}
\end{equation}
gives the full momentum dependence of the dimer-atom scattering function.

The integral equation shown in Eq.~(\ref{eqn:ieqn}) as well as the scattering 
function expression shown in Eq.~(\ref{eqn:a-scattering}) reduce to the known 
results for the equal-mass particles case~\cite{brodsky, levinsen} 
when $m_A = m_B = m$. Since only the fermion exchange process is taken 
into account in the Born approximation, and that neglecting the second 
term on the right hand side of Eq.~(\ref{eqn:ieqn}) leads to 
$a_{DA} = 2 m_{DA} a_{AB}/m_{AB}$, which is consistent with the many-body 
results~\cite{iskin-mixture}. However, we need to include both terms 
and solve the integral equation in order to find the exact dimer-atom 
scattering length.

\subsection{Dimer-atom scattering function}
\label{sec:a_DA}

Next, we solve numerically the integral equation given in Eq.~(\ref{eqn:ieqn})
as a function of the mass ratio $m_A/m_B$ of the constituent particles 
of the dimer~\cite{recall}. For this purpose, it is convenient to change 
the upper integration limit to a finite value by a change of variables, 
e.g. $p a_{AB} = (1-x)/(1+x)$ and $q a_{AB} = (1-y)/(1+y)$ 
where $1 \ge \{x,y\} \ge -1$. The resultant integral is calculated 
by using the Gaussian-Legendre quadrature method, and using this 
discretization, we reduce the integral equation to a matrix-eigenvalue problem.

\begin{figure} [htb]
\centerline{\scalebox{0.6}{\includegraphics{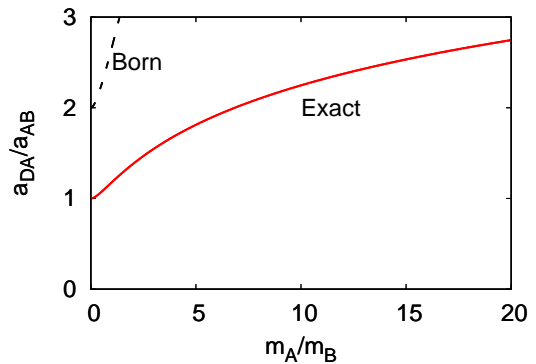}}}
\caption{\label{fig:a_DA}
The exact scattering length $a_{DA} = a_0^{DA}(0)$ between an
AB dimer (dimer D consists of one fermionic particle A and a second 
particle B) and a fermionic particle A is shown as a function of mass 
ratio $m_A/m_B$ (solid red line)~\cite{recall}. Particle B could be a
different species of fermion or a boson. Note that the disagreement
between the exact value and the Born approximation one (dashed line) 
increases rapidly as the mass ratio increases. 
}
\end{figure}

The exact solutions and the Born approximation values of $a_{DA} = a_0^{DA}(0)$ 
are shown in Fig.~\ref{fig:a_DA}. When $m_A = m_B$, we find 
$a_{DA} \approx 1.18a_{AB}$, which is in agreement with the results previously 
found for equal-mass particles~\cite{skorniakov, petrov-abf, brodsky, levinsen}.
The scattering length $a_{DA}$ increases (decreases) from this value 
with increasing (decreasing) mass ratio, and $a_{DA} \to a_{AB}$ 
in the limit of $m_A/m_B \to 0$ as expected. These results exactly match 
the few-body results of Petrov~\cite{petrov-abf}. It is quite remarkable that
the diagrammatic $T$-matrix approach exactly recovers the few-body results 
for all mass ratios, since the diagrammatic approach is performed in 
momentum space, while the few body approach is performed in real 
space~\cite{note}. Note that the Born approximation values for $a_{DA}$ 
are not in agreement with the exact values for any mass ratio, 
and that the disagreement increases rapidly with increasing mass ratio, 
but the general qualitative trend is captured by the Born approximation 
as can be seen in Fig.~\ref{fig:a_DA}.

\begin{figure} [htb]
\centerline{\scalebox{0.6}{\includegraphics{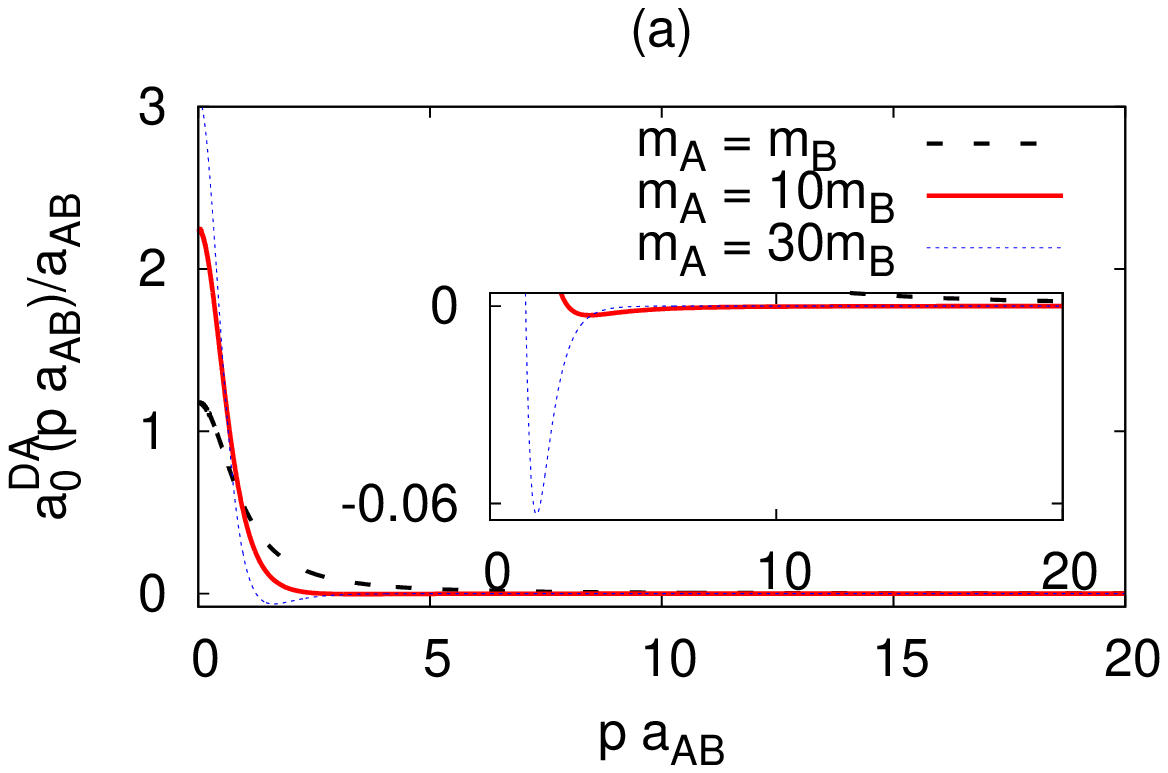}}}
\centerline{\scalebox{0.6}{\includegraphics{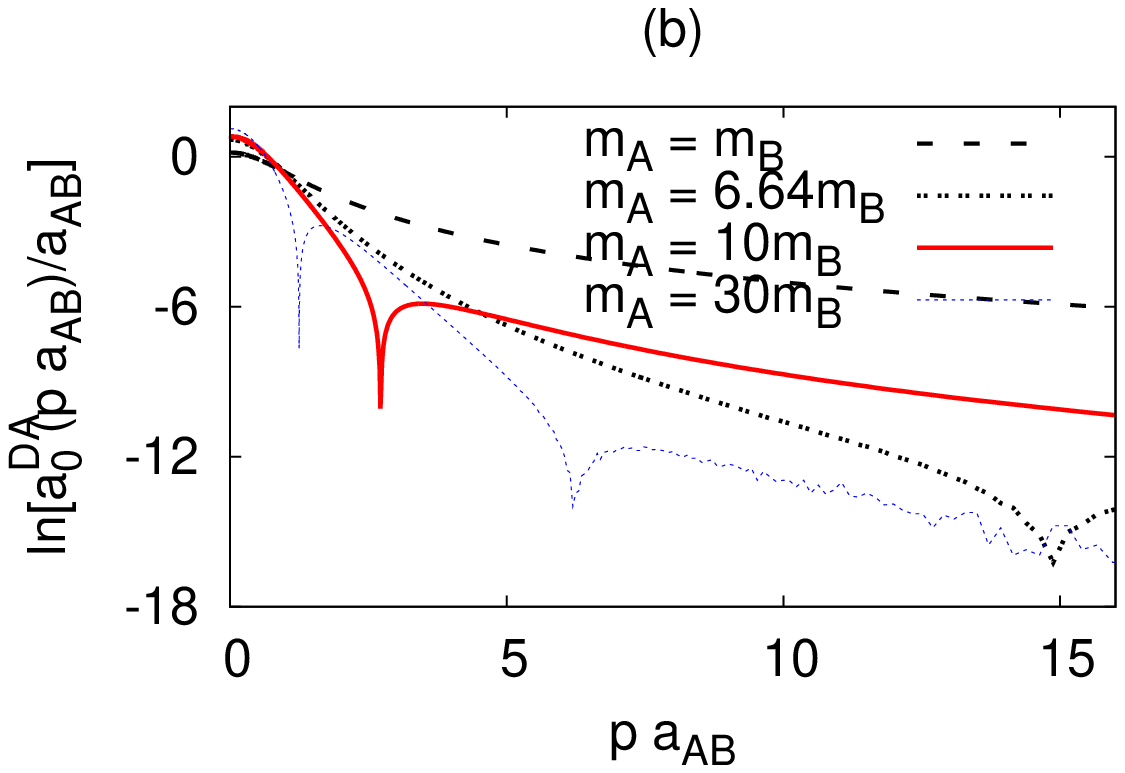}}}
\caption{\label{fig:a_DA_k}
The scattering function $a_0^{DA}(p)$ between an
AB dimer (dimer D consists of one fermionic particle A and a second 
particle B) and a fermionic particle A is shown as a function of the
outgoing momentum $p$ for some mass ratios $m_A/m_B$~\cite{recall}. 
Particle B could be a different species of fermion or a boson. 
Here, all length scales are in units of the two-body scattering length
between A and B particles $a_{AB}$, and in (b) we show $\ln[a_0^{DA}(p)]$ .
}
\end{figure}

In Fig.~\ref{fig:a_DA_k}, we show the scattering function $a_0^{DA}(p)$ 
as a function of the outgoing momentum $p$ for some mass ratios $m_A/m_B$. 
When $m_A = m_B$, this is a monotonically decreasing (positive)
function of $p$ with a long tail. However, beyond some critical 
mass ratio, this behavior changes dramatically. For instance, 
when $m_A = 10m_B$, after an initial rapid drop, 
$a_0^{DA}(p)$ function changes sign and becomes negative 
at $p a_{AB} \approx 2.72$. Beyond this critical momentum, 
it reaches a minimum value and slowly decays to zero from the 
negative side. As the mass ratio gets lower (higher), $a_0^{DA}(p)$ 
changes sign at higher (lower) momenta. For instance, 
when $m_A = 6.64m_B$, corresponding to a three-body system consisting 
of one $^6$Li atom and two $^{40}$K atoms, it changes sign at least 
once at $p a_{AB} \approx 14.7$ (but it changes sign at least 
once at $p a_{AB} \approx 1.96$ when $m_A = 13.61m_B$).
We find clear sign changes when $m_A \gtrsim 6.5m_B$, but we donot 
know whether $a_0^{DA}(p)$ changes sign for even lower mass ratios 
since precision issues obscure the results.

Although our calculation is not reliable for mass ratios above 
$13.61$~\cite{recall}, for illustration purposes, in 
Fig.~\ref{fig:a_DA_k}(b), we show $\ln[a_0^{DA}(p)]$ as a function 
of $p$ when $m_A = 30m_B$. In this case, the scattering function 
first changes sign and becomes negative at $p a_{AB} \approx 1.25$, 
but then it changes sign again and becomes positive at 
$p a_{AB} \approx 6.2$. Since the large-$p$ behavior of this 
function is again obscured due to precision issues, as can be seen 
in the figure, we could not resolve whether it has more sign 
changes at larger $p$ values. However, we mention that as the mass 
ratio increases further to $100$, $a_0^{DA}(p)$ changes sign at 
least 4 times, and the oscillating pattern becomes apparent.

This effect could be related to the ``fall of a particle to the 
center'' phenomenon~\cite{landau, petrov}, and to the 
emergence of three-body Efimov bound states~\cite{efimov, kartavtsev},
which are known to occur when the mass ratio is large.
When the heavy A particles are separated 
from each other by a distance $r \ll a_{AB}$ (or equivalently 
$p a_{AB} \gg 1$), it is known that an exchange of a light B 
particle mediates an effective $-C/(m_A r^2)$ attraction 
between A particles. Here, the coefficient $C \approx 0.162m_A/m_B$ 
increases with increasing mass ratio~\cite{efimov}. 
Since A particles are fermions, this attraction 
competes with the Pauli repulsion which manifests itself as a 
centrifugal $\ell(\ell+1)/(m_A r^2)$ potential, where 
$\ell$ is the angular momentum. Therefore, when the masses 
are comparable ($C \approx 0.162$), the Pauli repulsion is 
about one order of magnitude stronger than the mediated 
attraction. When the mass ratio $m_A/m_B$ reaches a critical
value, i.e. when $C = \ell(\ell+1)$, the effective 
interaction (mediated attraction plus centrifugal repulsion) 
between A particles vanishes. For the lowest p-wave angular 
momentum channel $\ell = 1$, this occurs 
when $m_A \gtrsim 12.33m_B$.

Beyond this critical mass ratio, there is a second critical
mass ratio beyond which the effective attraction between A 
particles is strong enough that the A particles stay in an 
infinitely small region around each other, i.e. the remaining 
A particle falls to the center of the attraction~\cite{landau}.
This second critical mass ratio can be approximated by
$\gamma_c = 1/4 = C - \ell(\ell+1)$~\cite{landau2}.
For the lowest p-wave angular momentum channel $\ell = 1$, 
this occurs when $m_A \gtrsim 13.85m_B$. Note that, since the 
B particle is already bound to one of the A particles, 
``fall of a particle to the center'' is related to emergence 
of three-body Efimov bound states. More accurate calculations 
show that the latter occurs when 
$m_A \gtrsim 13.61m_B$~\cite{kartavtsev, petrov}.

Having presented the diagrammatic $T$-matrix approach for the 
three-body (s-wave) dimer-atom scattering, for all mass 
ratios $m_A/m_B$, next we discuss the relevance of our results 
to the quantum phases of two-species Fermi-Fermi and Bose-Fermi 
mixtures of atomic gases at ultracold temperatures.

\subsection{Ultracold Fermi-Fermi and Bose-Fermi mixtures}
\label{sec:mixtures}

It has been shown that~\cite{pieri06, iskin-mixture} (see~\cite{shin08} 
for experimental confirmation), in the strong-attraction 
or molecular limit, two-species Fermi-Fermi mixtures with 
population imbalance can be well-described by effective Bose-Fermi 
models, where fermion-fermion pairs behave as molecular 
bosons (dimers) and interact weakly with each other and with 
the remaining unpaired (excess) fermions. 
These simpler models only require accurate scattering lengths 
between two molecular bosons (dimer-dimer, i.e. $a_{DD}$), and between a 
molecular boson and an unpaired fermion (dimer-atom, i.e. $a_{DA}$). 
Note that the exact three- and four-body results in vacuum 
are sufficient to describe ultracold atomic mixtures in the 
molecular limit, since experiments are always performed at low densities. 
Several fermionic atoms ($^{6}$Li, $^{40}$K, $^{87}$Sr~\cite{innsbruck}, 
and $^{171}$Yb~\cite{fukuhara}) are being currently investigated,
and experimental methods for studying two-species Fermi-Fermi 
mixtures are being developed in several groups, e.g.$^{6}$Li-$^{40}$K 
mixture~\cite{taglieber08,wille08,voigt09,spiegelhalder09,tiecke09,spiegelhalder10}. 
Thus, anticipating future experiments involving various other mixtures, 
in Table~\ref{table:ff}, we list the exact dimer-atom scattering lengths 
for all possible mixtures.

\begin{table} [htb]
\begin{tabular}{|c||c|c|c|c|c|}
\hline
 B $\setminus$ A 	 & $^6$Li & $^{40}$K & $^{87}$Sr & $^{171}$Yb & $^{173}$Yb \\
\hline
\hline
$^{6}$Li & 1.17907 & 1.98106 & 2.50583 & 3.01531 & 3.02437 \\
	 & (1.00000) & (6.64392) & (14.4484) & (28.4178) & (28.7506) \\
\hline
$^{40}$K & 1.01033 & 1.17907 & 1.41148 & 1.72559 & 1.73187 \\
	 & (0.15051) & (1.00000) & (2.17468) & (4.27726) & (4.32735) \\
\hline
$^{87}$Sr & 1.00251 & 1.06337 & 1.17907 & 1.37374 & 1.37799 \\
	 & (0.06921) & (0.45984) & (1.00000) & (1.96685) & (1.98988) \\
\hline
$^{171}$Yb & 1.00069 & 1.02195 & 1.07331 & 1.17907 & 1.18159 \\
	 & (0.03519) & (0.23379) & (0.50843) & (1.00000) & (1.01171) \\
\hline
$^{173}$Yb & 1.00067 & 1.02153 & 1.07209 & 1.17657 & 1.17907 \\
	 & (0.03478) & (0.23109) & (0.50254) & (0.98842) & (1.00000) \\
\hline
\end{tabular}
\caption{\label{table:ff} 
The exact scattering length $a_{DA} = a_0^{DA}(0)$ between a bosonic 
AB dimer (consisting of one fermionic A atom and one fermionic B atom) 
and a fermionic A atom is shown for a list of two-species Fermi-Fermi 
mixtures. Here, $a_{DA}$ is in units of the two-body scattering length 
between A and B particles $a_{AB}$, and the mass ratios $m_A/m_B$ are 
shown inside the parenthesis~\cite{recall}.
}
\end{table}

Similarly, in the strong-attraction or molecular limit, two-species 
Bose-Fermi mixtures (with more fermions than bosons, otherwise 
see~\cite{petrov10}) can be well-described by effective Fermi-Fermi 
models, where boson-fermion pairs can be shown to behave as 
molecular fermions (dimers) and interact weakly with the 
remaining unpaired fermions. These simpler models again only require 
accurate scattering length between a molecular fermion and an unpaired 
fermion (dimer-atom, i.e. $a_{DA}$). Experimental methods for studying 
two-species Bose-Fermi mixtures are also being developed in several 
groups~\cite{inouye04, ospelkaus06, zaccanti, zirbel}, 
and in Table~\ref{table:bf}, we list the exact dimer-atom scattering 
lengths for all possible mixtures.

\begin{table} [htb]
\begin{tabular}{|c||c|c|c|c|c|}
\hline
 B $\setminus$ A	& $^6$Li & $^{40}$K & $^{87}$Sr & $^{171}$Yb & $^{173}$Yb \\
\hline
\hline
$^{7}$Li & 1.14817 & 1.88773 & 2.39579 & 2.89621 & 2.90516 \\
	 & (0.85734) & (5.69612) & (12.3872) & (24.3638) & (24.6491) \\
\hline
$^{23}$Na & 1.026642 & 1.33055 & 1.66074 & 2.05182 & 2.05926 \\
	 & (0.26164) & (1.73834) & (3.78033) & (7.43532) & (7.52240) \\
\hline
$^{39}$K & 1.01080 & 1.18459 & 1.42138 & 1.73929 & 1.74563 \\
	 & (0.15438) & (1.02567) & (2.23051) & (4.38707) & (4.43844) \\
\hline
$^{41}$K & 1.00988 & 1.17371 & 1.40181 & 1.71213 & 1.71835 \\
	 & (0.14677) & (0.97516) & (2.12067) & (4.17103) & (4.21988) \\
\hline
$^{84}$Sr & 1.00268 & 1.06670 & 1.18675 & 1.38665 & 1.39100 \\
	 & (0.07168) & (0.47625) & (1.03570) & (2.03706) & (2.06091) \\
\hline
$^{85}$Rb & 1.00262 & 1.06556 & 1.18413 & 1.38226 & 1.38658 \\
	 & (0.07084) & (0.47065) & (1.02352) & (2.01310) & (2.03668) \\
\hline
$^{86}$Sr & 1.00256 & 1.06445 & 1.18157 & 1.37796 & 1.38225 \\
	 & (0.07002) & (0.46519) & (1.01163) & (1.98973) & (2.01303) \\
\hline
$^{87}$Rb & 1.00251 & 1.06337 & 1.17906 & 1.37373 & 1.37799 \\
	 & (0.06921) & (0.45984) & (0.99999) & (1.96684) & (1.98987) \\
\hline
$^{88}$Sr & 1.00245 & 1.06232 & 1.17662 & 1.36961 & 1.37382 \\
	 & (0.06843) & (0.45462) & (0.98866) & (1.94454) & (1.96732) \\
\hline
$^{133}$Cs & 1.00112 & 1.03306 & 1.10408 & 1.23978 & 1.24291 \\
	 & (0.04526) & (0.30069) & (0.65391) & (1.28615) & (1.30121) \\
\hline
$^{135}$Cs & 1.00109 & 1.03228 & 1.10200 & 1.23581 & 1.23890 \\
	 & (0.04459) & (0.29624) & (0.64422) & (1.26708) & (1.28192) \\
\hline
$^{168}$Yb & 1.00071 & 1.02260 & 1.07519 & 1.18292 & 1.18548 \\
	 & (0.03582) & (0.23797) & (0.51752) & (1.01758) & (1.02980) \\
\hline
$^{170}$Yb & 1.00070 & 1.02217 & 1.07393 & 1.18034 & 1.18297 \\
	 & (0.03540) & (0.23517) & (0.51142) & (1.00589) & (1.01767) \\
\hline
$^{172}$Yb & 1.00068 & 1.02174 & 1.07270 & 1.17781 & 1.18032 \\
	 & (0.03498) & (0.23243) & (0.50547) & (0.99418) & (1.00583) \\
\hline
$^{174}$Yb & 1.00067 & 1.02132 & 1.07150 & 1.17534 & 1.17783 \\
	 & (0.03458) & (0.22976) & (0.49965) & (0.98274) & (0.99425) \\
\hline
$^{176}$Yb & 1.00065 & 1.02092 & 1.07033 & 1.17292 & 1.17538 \\
	 & (0.03419) & (0.22714) & (0.49396) & (0.97155) & (0.98292) \\
\hline
\hline
\end{tabular}
\caption{\label{table:bf} 
The exact scattering length $a_{DA} = a_0^{DA}(0)$ between a fermionic 
AB dimer (consisting of one fermionic A atom and one bosonic B atom) 
and a fermionic A atom is shown for a list of two-species Bose-Fermi 
mixtures. Here, $a_{DA}$ is in units of the two-body scattering length 
between A and B particles $a_{AB}$, and the mass ratios $m_A/m_B$ are shown 
inside the parenthesis~\cite{recall}.
}
\end{table}

As we pointed out above, the three-body scattering problem is universal
for mass ratios below $13.61$, and therefore, the dimer-atom scattering is 
proportional to the two-body scattering length $a_{AB}$ with the 
proportionality factor depending only on the masses of the constituent 
particles. However, since three-body Efimov bound states emerges
for larger mass ratios~\cite{efimov, kartavtsev}, 
this problem is not universal (an additional parameter coming from the 
short-range three-body physics is needed for an accurate description),
and our analysis does not include this non-universal effect~\cite{petrov-abf, petrov}. 
In particular, the $^6$Li-$^{87}$Sr, $^6$Li-$^{171}$Yb and 
$^6$Li-$^{173}$Yb Fermi-Fermi mixtures and $^7$Li-$^{171}$Yb,
and $^7$Li-$^{173}$Yb Bose-Fermi mixtures have mass ratios that 
are above the critical ratio for the emergence of three-body 
Efimov bound states.

\section{Conclusions}
\label{sec:conclusions}

In this paper, we used the diagrammatic $T$-matrix approach to analyze 
the three-body scattering problem between two identical fermions and 
a third particle. The third particle could be a different species of 
fermion or a boson. We calculated the exact s-wave dimer-atom scattering 
length for all mass ratios, and our results exactly match the few-body 
results of Petrov who obtained them by solving the three-body 
Schr\"odinger equation~\cite{petrov-abf}. It is quite remarkable that the 
diagrammatic $T$-matrix approach exactly recovers the few-body results 
for all mass ratios, since the diagrammatic approach is performed in 
momentum space, while the few-body approach is performed in real 
space~\cite{note}.

We also discussed the relevance of our results to the quantum 
phases of two-species Fermi-Fermi and Bose-Fermi mixtures of 
atomic gases at ultracold temperatures. In particular, in the 
strong-attraction or molecular limit, these mixtures can be 
well-described by simpler effective models, where paired atoms 
behave as dimers and interact weakly with each other and with 
the remaining unpaired atoms. These effective descriptions 
require only the scattering lengths between two dimers, and 
between a dimer and an unpaired atom. Anticipating future 
experiments, we listed the exact dimer-atom scattering lengths for 
all available two-species Fermi-Fermi and Bose-Fermi mixtures. 

In addition, we showed that, unlike that of the equal-mass particles 
case where the three-body scattering $T$-matrix decays monotonically 
as a function of the outgoing momentum, after an initial rapid 
drop, this function changes sign and becomes negative 
at large momenta and then decays slowly to zero when the mass 
ratio of the fermions to the third particle is higher than 
a critical value (around $6.5$). As the mass ratio gets higher, 
modulations of the $T$-matrix become more apparent with 
multiple sign changes. We argued that this effect
could be related to the ``fall of a particle to the center'' 
phenomenon~\cite{landau, petrov} and to the emergence of three-body 
Efimov bound states~\cite{efimov, kartavtsev}.

\section{Acknowledgments}
\label{sec:ack}

The author thanks P. Pieri for useful discussions, and The Scientific 
and Technological Research Council of Turkey (T\"{U}B$\dot{\mathrm{I}}$TAK) 
for financial support.


\begin{thebibliography}{99}
\bibitem{skorniakov} G. V. Skorniakov and K. A. Ter-Martirosian, Zh. Eksp. Teor. Fiz. \textbf{31}, 775 (1956); and Sov. Phys. JETP \textbf{4}, 648 (1957).
\bibitem{brodsky} I. V. Brodsky, M. Yu. Kagan, A. V. Klaptsov, R. Combescot, and X. Leyronas, Phys. Rev. A \textbf{73}, 032724 (2006).
\bibitem{levinsen} J. Levinsen and V. Gurarie, Phys. Rev. A \textbf{73}, 053607 (2006).
\bibitem{bedaque98} P. F. Bedaque and U. van Kolck, Physics Letters B \textbf{428}, 221 (1998).
\bibitem{petrov-abf} D. S. Petrov, Phys. Rev. A \textbf{67}, 010703(R) (2003).
\bibitem{iskin-bf} M. Iskin and C. A. R. S{\'a} de Melo, Phys. Rev. A \textbf{77}, 013625 (2008).
\bibitem{efimov} V. N. Efimov, Yad. Fiz. \textbf{12}, 1080 (1970); Sov. J. Nucl. Phys. \textbf{12}, 589 (1971); Nucl. Phys. A \textbf{210}, 157 (1973).
\bibitem{kartavtsev} O. I. Kartavtsev and A. V. Malykh, J. Phys. B \textbf{40}, 1429 (2007).
\bibitem{petrov} D. S. Petrov, C. Salomon, and G. V. Shlyapnikov, J. Phys. B: At. Mol. Opt. Phys. \textbf{38}, S645 (2005).

\bibitem{taglieber08} M. Taglieber, A.-C. Voigt, T. Aoki, T. W. H\"ansch, and K. Dieckmann, Phys. Rev. Lett. \textbf{100}, 010401 (2008).
\bibitem{wille08} E. Wille, F. M. Spiegelhalder, G. Kerner, D. Naik, A. Trenkwalder, G. Hendl, F. Schreck, R. Grimm, T. G. Tiecke, J. T. M. Walraven, S. J. J. M. F. Kokkelmans, E. Tiesinga, and P. S. Julienne, Phys. Rev. Lett. \textbf{100}, 053201 (2008).
\bibitem{voigt09} A.-C. Voigt, M. Taglieber, L. Costa, T. Aoki, W. Wieser, T. W. H\"ansch, and K. Dieckmann,  Phys. Rev. Lett. \textbf{102}, 020405 (2009).
\bibitem{spiegelhalder09} F. M. Spiegelhalder, A. Trenkwalder, D. Naik, G. Hendl, F. Schreck, and R. Grimm, Phys. Rev. Lett. \textbf{103}, 223203 (2009).
\bibitem{tiecke09} T. G. Tiecke \textit{et al.}, arXiv:0908.2071 (2009).
\bibitem{spiegelhalder10} F. M. Spiegelhalder \textit{et al.}, arXiv:1001.5253 (2010).

\bibitem{inouye04} S. Inouye, J. Goldwin, M. L. Olsen, C. Ticknor, J. L. Bohn, and D. S. Jin,  Phys. Rev. Lett. \textbf{93}, 183201 (2004).
\bibitem{ospelkaus06} S. Ospelkaus, C. Ospelkaus, L. Humbert, K. Sengstock, and K. Bongs,  Phys. Rev. Lett. \textbf{97}, 120403 (2006).
\bibitem{zaccanti} M. Zaccanti, C. D'Errico, F. Ferlaino, G. Roati, M. Inguscio, and G. Modugno, Phys. Rev. A \textbf{74}, 041605(R) (2006).
\bibitem{zirbel} J. J. Zirbel, K.-K. Ni, S. Ospelkaus, J. P. D’Incao, C. E. Wieman, J. Ye, and D. S. Jin,  Phys. Rev. Lett. \textbf{100}, 143201 (2008).

\bibitem{iskin-mixture} M. Iskin and C. A. R. S{\'a} de Melo, Phys. Rev. Lett. \textbf{97}, 100404 (2006); Phys. Rev. A \textbf{76}, 013601 (2007).
\bibitem{recall} Our results are accurate only for mass ratios that are smaller than the critical value $13.61$, beyond which an additional parameter coming from the short-range (or large-momentum) three-body physics is needed for an accurate description~\cite{efimov, petrov-abf, petrov, kartavtsev}, which is beyond the scope of this paper.
\bibitem{note} Two methods are equivalent because all of the possible scattering processes are taken into account exactly in our diagrammatic approach at zero temperature.
\bibitem{landau} L. D. Landau and E. M. Lifshitz, \textit{Quantum Mechanics}, pp. 114 (Butterworth-Heinemann, Oxford, 1999).
\bibitem{landau2} See Eq.~(35.2) and the discussion above Eq.~(35.10) for the two-body scattering problem in~\cite{landau}.
\bibitem{pieri06} P. Pieri and G. C. Strinati, Phys. Rev. Lett. \textbf{96}, 150404 (2006).
\bibitem{shin08} Yong-il Shin, Andr\'e Schirotzek, Christian H. Schunck, and Wolfgang Ketterle, Phys. Rev. Lett. \textbf{101}, 070404 (2008).
\bibitem{innsbruck} R. Grimm, private communication.
\bibitem{fukuhara} T. Fukuhara, Y. Takasu, M. Kumakura, and Y. Takahashi, Phys. Rev. Lett. \textbf{98}, 030401 (2007).
\bibitem{petrov10} K. Helfrich, H.-W. Hammer, and D.S. Petrov, arXiv:1001.4371 (2010).
\end{thebibliography}
\end{document}